\documentclass[fleqn]{2023SCGE}
\UseRawInputEncoding
\newcommand{\xiamen}{Department of Physics and Key Laboratory of Low Dimensional
Condensed Matter Physics (Department of Education of Fujian Province), \\Xiamen
University, Xiamen 361005, Fujian, China}
\newcommand{\lanzhou}{Lanzhou Center for Theoretical Physics, Lanzhou University,
Lanzhou 730000, Gansu, China}

\usepackage{ulem}
\begin{document}

\ensubject{subject}

\ArticleType{Article}
\SpecialTopic{SPECIAL TOPIC: }
\Year{Year}
\Month{Month}
\Vol{66}
\No{1}
\DOI{??}
\ArtNo{000000}
\ReceiveDate{Month Day, Year}
\AcceptDate{Month Day, Year}

\title{A pseudoclassical theory for the wavepacket dynamics of the kicked rotor model}{A pseudoclassical theory for the wavepacket dynamics of the kicked rotor model}

\author[1]{Zhixing Zou}{}%
\author[1, 2]{Jiao Wang}{{phywangj@xmu.edu.cn}}

\AuthorMark{Zou Z X}

\AuthorCitation{Zou Z X, Wang J}

\address[1]{\xiamen}
\address[2]{\lanzhou}


\abstract{In this study, we propose a generalized pseudoclassical theory for the kicked rotor model in an attempt to discern the footprints of the classical dynamics in the deep quantum regime. Compared with the previous pseudoclassical theory that applies only in the neighborhoods of the lowest two quantum resonances, the proposed theory is applicable in the neighborhoods of all quantum resonances in principle by considering the quantum effect of the free rotation at a quantum resonance. In particular, it is confirmed by simulations that the quantum wavepacket dynamics can be successfully forecasted based on the generalized pseudoclassical dynamics, offering an intriguing example where it is feasible to bridge the dynamics in the deep quantum regime to the classical dynamics. The application of the generalized pseudoclassical theory to the $\mathcal{PT}$-symmetric kicked rotor is also discussed.}



\maketitle


\begin{multicols}{2}
\section{Introduction}

Quantum-classical correspondence in a general system is the core issue of quantum chaos research or quantum chaology~\cite{Berry89}. Based on the quantum-classical correspondence principle, the characteristics of a classically chaotic system are anticipated to manifest themselves in the corresponding quantum system in the semiclassical limit $\hbar\to 0$ ($\hbar$ is the effective Planck constant). Indeed, after 4 decades of intensive research, the general quantum manifestations of classical chaos, such as spectral statistics and the morphologies of wavefunctions, have been well revealed~\cite{Haake2018}. However, for quantum systems in the deep quantum regime opposite to the semiclassical limit, thus far, the relation\Authorfootnote

\noindent  between them and their classical counterparts have not been systematically addressed yet. The question arises: Are there any close connections between the classical chaotic dynamics and the quantum properties in the deep quantum regime? Without the guide of any physical law or principle, the answer is not immediate.

An interesting and illuminating example is the pseudoclassical dynamics~\cite{Fishman2002, Fishman2003} found in the kicked rotor model~\cite{Casati79}, a paradigm of quantum chaos with a free rotor subjected to periodic stroboscopic external kicks. Its properties significantly depend on the dimensionless parameter $\alpha=\hbar T/I$, where $T$ is the kicking period and $I$ is the rotational inertia of the rotor. For a general value of $\alpha$, the quantum kicked rotor follows the classical diffusive dynamics first and then the quantum dynamical localization eventually takes over ~\cite{Casati79, Casati95}; however, by contrast, for $\alpha$ being a rational multiple of $4\pi$, that is, $\alpha=4\pi r/s$ with coprime integers $r$ and $s$, a quantum state usually ballistically spreads, which is named as ``quantum resonance''~\cite{Casati95, Izrailev1980}. For low-order resonance with a small $s$ value, the system is in the deep quantum regime; however, surprisingly, Fishman {\it et al.}~\cite{Fishman2002, Fishman2003} observed that when $\alpha$ is slightly detuned from the resonance condition by a nonzero $\delta$, the quantum motion can be interpreted according to a certain fictitious classical system that is different from but closely related to the original classical counterpart. The fictitious classical system is named the pseudoclassical system, and the limit $\delta \to 0$ is the pseudoclassical limit.  This result demonstrates a novel and unconventional aspect of quantum-classical correspondence.

Unfortunately, the so-developed pseudoclassical theory only directly applies near the lowest two resonances, that is, $s = 1$ and $2$. When the neighborhood of high-order resonance is considered, because a global pseudoclassical phase-space approximation regarding a unique classical Hamiltonian is impossible, one has to resort to local pseudoclassical approximations of different Hamiltonians related to the quasienergy bands of the considered quantum resonance~\cite{Rebuzzini2009}. Nevertheless, the interaction between the quasienergy bands may ruin the local pseudoclassical approximations to make their predictions invalid~\cite{Wang2013}. Recently~\cite{Zou2022}, we realized that owing to the periodicity of the phase space, in a system with a spherical or cylindrical phase space, the free rotation of a wavepacket at quantum resonance may lead to the simultaneous presence of multiple wavepackets, in clear contrast to the usual scenario where only one wavepacket is found throughout. Such a wavepacket-multiplying effect is a pure quantum interference effect and has not been considered earlier. Considering this, the pseudoclassical theory may find wide applications when being generalized by considering this effect. This conjecture has been confirmed with the kicked top model~\cite{Zou2022}, another paradigm of quantum chaos with a spherical phase space, which is closely connected with the kicked rotor~\cite{Haake1987, Haake1988}.

The objective of this work is to establish a generalized pseudoclassical theory for the kicked rotor model with a cylindrical phase space with regard to its significant role in quantum chaos research. This attempt is successful again. Based on our theory, for the lowest two resonances of $s = 1$ and $2$, the free rotation does not lead to the wavepacket multiplying, and this is why the previous pseudoclassical theory is valid in these two cases. However, for higher-order resonance, particularly that of an odd $s$, the wavepacket multiplying can be shown to occur for certain, and the generalized theory is thus imperative. Our study might be a preliminary but positive attempt to address the quantum-classical correspondence issue in the deep quantum regime.

This article is organized as follows: Sec. II briefly describes the kicked rotor model and discusses in detail the pseudoclassical limit of the wavepacket dynamics. Sec. III compares the quantum evolution of a wavepacket with the prediction of the pseudoclassical theory via numerical simulations, convincingly demonstrating that the latter is effective. Sec. IV presents the extension to the $\mathcal{PT}$-symmetric kicked rotor and confirmation by numerical simulations. Finally, Sec. V concludes this work.

\section{The pseudoclassical limit of the kicked rotor }

The Hamiltonian of the kicked rotor is
\begin{eqnarray}\label{H0}
H=\frac{P^2}{2I}+\frac{KI}{\omega T}\cos (\omega\theta)\sum_{m=-\infty}^{\infty} \delta(\tau-mT).
\end{eqnarray}
Here, $I$, $P$, and $\theta$ are the rotational inertia, the angular momentum, and the conjugate angular coordinate of the rotor, respectively, whereas $T$ and $K$ are the kicking period and the dimensionless kicking strength, respectively. The integer parameter $\omega$ is introduced for our aim here, which is unity in the conventional kicked rotor model. The angular coordinate $\theta$ is imposed with a period of $2\pi$, such that the phase space is a cylinder.

For the quantum kicked rotor, it is convenient to adopt the basis of the angular momentum eigenstates, $\{|n\rangle; -\infty<n<\infty\}$, where $n$ is an integer and $|n\rangle$ meets $P|n\rangle =n\hbar|n\rangle$. Because the Hamiltonian is periodic with time $T$, the evolution of the rotor for time $T$ can be fulfilled by applying the Floquet operator
\begin{eqnarray}\label{U0}
U=\exp\left(-i\frac{\alpha}{2}\hat \nu ^2\right)\exp(-i\frac{K}{\alpha\omega}\cos (\omega\theta))
\end{eqnarray}
to its current state just before a kick, where $\alpha\equiv \hbar T/I$ and $\hat \nu \equiv P/\hbar$. As stated in the Introduction, the quantum dynamics depends qualitatively on the fact if $\alpha$ is a rational multiple of $4\pi$. The case $\alpha = 4\pi r/s$ with coprime integers $r$ and $s$ corresponds to quantum resonance, and the quantum state usually ballistically spreads. An exception is for $s = 2$ and $\omega$ being odd so that $U^2 = 1$, suggesting that the quantum state does not change after time $2T$, which is termed as ``quantum antiresonance''~\cite{Izrailev1980, Dana1996}.

When the system is slightly detuned away from a quantum resonance, a pseudoclassical theory has been developed to address the quantum dynamics via a classical map, the so-called pseudoclassical limit~\cite{Fishman2002, Fishman2003}. However, this theory only works at the lowest two resonances. In the following of this section, we attempt to extend this theory to the neighborhoods of higher-order resonances with $\alpha=4\pi r/s +\delta$, where $\delta$ (incommensurate to $\pi$) is a weak perturbation to the resonance condition. To perform a close comparison between the quantum and the classical dynamics, which is critical to our objective, we invoke the (squeezed) coherent state in the former, with regard to the advantages that the Husimi distribution of a coherent state has the minimum uncertainty in the phase space and its center point in the phase space represents exactly the classical counterpart of the coherent state. For convenience explained later, we use $(p, \theta)$ to denote a classical state, or a point in the phase space, with $p\equiv P(T/I)(\delta/\alpha)$. The expression of the coherent state centered at $(p,\theta)$ is
\begin{eqnarray}\label{coherent}
|p,\theta\rangle = c\sum_{n}\exp\left(-\frac{\delta}{2}(n-{\frac{p}{\delta}})^2\right)
\exp\left(-in\theta \right)|n\rangle,
\end{eqnarray}
where $c$ is the normalization factor. When $\delta$ is small, the uncertainty of $p$ and $\theta$ is the same, $\delta_{p}\approx \delta_{\theta}\approx\sqrt{\delta/2}$. In the phase space, the Husimi distribution of the coherent state is a Gaussian function centered at $(p,\theta)$. In the limit of $\delta \to 0$, it collapses to the point $(p,\theta)$.

Our task is to determine the one-step evolution for the classical state $(p,\theta)$ by analogy according to the corresponding quantum evolution for the coherent state $|p,\theta\rangle$. To this end, note that for $\alpha=4\pi r/s+\delta$, the Floquet operator can be rewritten as
\begin{eqnarray}\label{U}
U=\exp\left(-i2\pi\frac{r}{s}\hat \nu^2\right)\exp\left(-i\frac{\delta}{2}\hat \nu^2\right)
\exp(-i\frac{k}{\delta \omega}\cos\omega\theta),
\end{eqnarray}
where $k= K\delta/\alpha$. It includes two parts, i.e., $U=U_f U_\delta$, with $U_f=\exp(-i2\pi \frac{r}{s}\hat\nu^2)$ and $U_\delta =\exp \left(-i\frac{\delta }{2}\hat\nu^2\right)\exp(-i\frac{k}{\delta \omega}\cos(\omega\theta))$. The former, $U_f$, represents a pure free rotation, while the latter, $U_\delta$, represents kicked rotor dynamics (see Eq.~\eqref{U0}) with $\alpha\to \delta$ and $K\to k$. We thus break our task down into two steps. First, note that in the limit $\delta\to 0$, the quantum operation $U_\delta$ has a well-defined semiclassical limit. This can be observed more clearly by imagining $\delta$ as a virtual Planck constant so that we can write down the classical Hamiltonian corresponding to $U_\delta$ as
\begin{eqnarray}\label{Hdelta}
H_\delta=\frac{p^2}{2}+\frac{k}{\omega}\cos (\omega\theta)\sum_{n=-\infty}^{\infty}
\delta(\tau-n).
\end{eqnarray}
Based on $H_\delta$, the classical Poincar\'e map corresponding to $U_\delta$, denoted as $\mathcal{M}_\delta$, can be derived straightforwardly. To be concrete, in terms of $p$ and $\theta$, the map $\mathcal{M}_\delta$ that evolves the state $(p, \theta)$ to $(\tilde p, \tilde \theta)$ reads
\begin{eqnarray}\label{Md}
\mathcal{M}_\delta:
\begin{cases}
\tilde p=p+k\sin(\omega\theta),\\
\tilde \theta=\theta+\tilde p.
\end{cases}
\end{eqnarray}
Note that the phase space portrait created by $\mathcal{M}_\delta$ is periodic in both $p$ and $\theta$ of period $2\pi/\omega$.

Due to this close relation between the quantum operation $U_\delta$ and the classical map $\mathcal{M}_\delta$ in the limit of $\delta\to 0$, we assume that
\begin{eqnarray}\label{MQ}
|\tilde p,\tilde \theta\rangle = U_\delta |p,\theta\rangle
\end{eqnarray}
as well when $\delta$ is small, which is the only significant approximation we adopt for our theory. Equation \eqref{MQ} is the counterpart of the classical map $\mathcal{M}_\delta: (p, \theta)\to(\tilde p,\tilde \theta)$ (see Eq.~\eqref{Md}), and thus, we finish the first step of our task.

Afterward, we need to derive $U_f |\tilde p,\tilde \theta\rangle$ and work out its classical counterpart. The detailed calculation of $U_f |\tilde p,\tilde \theta\rangle$ is given in Appendix A, resulting in
\begin{eqnarray}\label{Uf1}
U_f \left|\tilde p,\tilde \theta\right\rangle=
\sum_{l=0}^{s-1} G_l |\tilde p,\tilde \theta+2\pi\frac{r}{s} l\rangle,
\end{eqnarray}
where $G_l$ is the Gaussian sum~\cite{Berndt1998}
\begin{eqnarray}\label{Gl}
G_l = \frac{1}{s} \sum_{m=0}^{s-1} \exp\left(-i2\pi\frac{r}{s}m(m-l)\right).
\end{eqnarray}
The physical meaning of Eq.~\eqref{Uf1} is evident: The intermediate coherent state $|\tilde p,\tilde \theta \rangle$ is mapped by $U_f$ into $s$ coherent states whose centers are located along the line of $p = \tilde p$ in the phase space. They are separated in $\theta$ by $2\pi\frac{r}{s}$ (and its multiples) from each other, and each coherent state has an associated complex amplitude given by a Gaussian sum. This is a peculiar characteristic of the quantum rotor at resonance. Note that not all of these $s$ coherent states exist necessarily. A coherent state disappears if the associated amplitude $G_l$ disappears. Suppose that there are $\mathcal{N}$ nonzero amplitudes, Eq.~\eqref{Uf1} can be rewritten as
\begin{eqnarray}\label{Uf2}
U_f|\tilde p, \tilde \theta\rangle
=\sum_{j=1}^\mathcal{N} A_j \left|\tilde p, \tilde \theta+\Delta_j\right\rangle.
\end{eqnarray}		
Here, for the $j$th component coherent state, its amplitude $A_j$ is a nonzero Gaussian sum $G_l$, and its position bias $\Delta_j$ is associated with the subscript of $G_l$ by $\Delta_j = 2\pi l r/s$ mod $2\pi$. Similarly, as in the limit $\delta\to 0$, a coherent state reduces to a point in phase space, Eq.~\eqref{Uf2} can be interpreted pseudoclassically. That is, the intermediate classical state $(\tilde p,\tilde \theta)$ is mapped by the pseudoclassical counterpart of $U_f$, denoted as $\mathcal{M}_f$, into a set of $\mathcal{N}$ states, and each of these is associated with a complex amplitude,
\begin{eqnarray}\label{Mf}
\mathcal{M}_f: (\tilde p, \tilde \theta)\to
\{[(\tilde p, \tilde \theta+\Delta_j); A_j], j=1,\cdots,\mathcal{N}\}.
\end{eqnarray}		
This concludes the second step of our task.

Therefore, formally, the pseudoclassical map corresponding to the quantum evolution $U|p, \theta\rangle$ that we are seeking for, denoted as $\mathcal{M}$, can be expressed as $\mathcal{M}=\mathcal{M}_f \mathcal{M}_\delta$, that is,
\begin{eqnarray}\label{M}
\mathcal{M}: (p,\theta)\to \{[(\tilde p, \tilde \theta+\Delta_j); A_j], j=1,\cdots,\mathcal{N}\}.
\end{eqnarray}	
The intermediate state $(\tilde p, \tilde \theta)$ is related to $(p, \theta)$ by the map $\mathcal{M_\delta}$ (see Eq.~\eqref{Md}). This is the core result that we have obtained. As shown in the next section, it does allow the prediction of the quantum dynamics in such a pseudoclassical way. Here, we emphasize that the amplitudes $\lbrace A_j\rbrace$ are crucial to this end. Specifically, $|A_j|^2$ has to be taken as the weight of the $j\emph{}$th state $(\tilde p, \tilde \theta+\Delta_j)$ that it is associated with to assess the expected value of a given observable. In addition, the phases encoded in these amplitudes have to be simultaneously considered to correctly trace the quantum evolution.

Some remarks are in order. First, for the lowest two resonances $s = 1$ and $s = 2$, $\mathcal{N} = 1$ and the amplitude of the only resultant state is unity. In addition, the map $\mathcal{M}$ reduces to that given by the original pseudoclassical theory~\cite{Fishman2002, Fishman2003}, which has a seemingly pure classical form. For
higher-order resonance, even though the map $\mathcal{M}$ is substantially more complex, there are no quantum operations and parameters that are explicitly involved. Despite this, the map $\mathcal{M}$ should be understood as a mix of the quantum and the classical dynamics according to our reasoning for deriving $\mathcal{M}$. It might be appropriate to regard the seemingly simple form for the lowest two resonances as a coincidence.

Second, for an odd $s$, all $s$ Gaussian sums are nonzero~\cite{Berndt1998} so that $\mathcal{N} = s$; for an even $s$, it can be shown that half of them must be zero so that $\mathcal{N} = s/2$ (see Appendix B). In either case, the main challenge for the implementation of the pseudoclassical map $\mathcal{M}$ lies in the rapid proliferation of the involved states when $\mathcal{N}\ge 2$. In general, their number exponentially increases, $\sim \mathcal{N}^t$, as the number of iterations $t$ (or the evolving time $\tau=tT$). This suggests why quantum dynamics is more complicated from such a novel perspective. Thus, for a general case, it can be hoped that in practice, our pseudoclassical theory only works in a short time to predict quantum evolution, which can be attributed to the intrinsic complexity of quantum dynamics.

Third, however, for the following two cases equipped with an $s$- or $s/2$-fold translational symmetry in the $\theta$ direction of the phase space, that is,\\
\indent C1) $s$ is odd and $\omega=s$\\
\indent C2) $s$ is even and $\omega=s/2$\\
the proliferation problem can be prevented because from the second iteration on ($t\ge 2$), two or more states can be mapped into one so that the total number of resultant states remains bounded by $s$ in case C1 and by $s/2$ in case C2. Appendix B provides a detailed discussion.

In fact, note that for a general case, $U_f$ and $U_\delta$ do not commute. However, for case C1, they commute (see Appendix C), demonstrating that the operator for the $t$ steps of iteration, $(U_f U_\delta)^t$, can be split into $(U_f U_\delta)^t=U_f^t U_\delta^t$. As the state proliferation is exclusively caused by $U_f^t=\exp(-i2\pi t \frac{r}{s} \nu^2)$, we can conclude that the number of final resultant coherent states cannot surpass $s$ for any time $t$. Moreover, the corresponding pseudoclassical map can be expressed as $\mathcal{M}^t=\mathcal{M}_f^t \mathcal{M}_\delta^t$, based on which the computing of the pseudoclassical evolution can be substantially simplified.

In case C2, $U_f$ and $U_\delta$ do not generally commute. However, $U'_f$ and $U'_\delta$ commute, where $U'_f = U_f \exp(i \pi r \nu)$ and $U'_\delta = \exp(-i\pi r \nu)U_\delta$ (see Appendix C). Consequently, $(U_f U_\delta)^t = {U'_f}^t {U'_\delta}^t$, which has similar implications to case C1.

\section{Verification of the pseudoclassical theory}

The effectiveness of the pseudoclassical theory is checked by the comparison of its predictions on the wavepacket dynamics and that directly obtained with the quantum Floquet operator. Three representative examples are examined, of which one is for a general case with $r = 1$, $s = 4$, and $\omega = 1$, where the number of coherent states doubles after every step of iteration. The other two instances, one with $r = 1$, $s = 3$, and $\omega = 3$ and another with $r = 1$, $s = 4$, and $\omega = 2$, are for cases C1 and C2, respectively, where the proliferation of the coherent states is suppressed. As shown in the following of this section, the pseudoclassical theory effectively works in all three cases.

\begin{figure}[H]
\centering
\includegraphics[width=7.9cm]{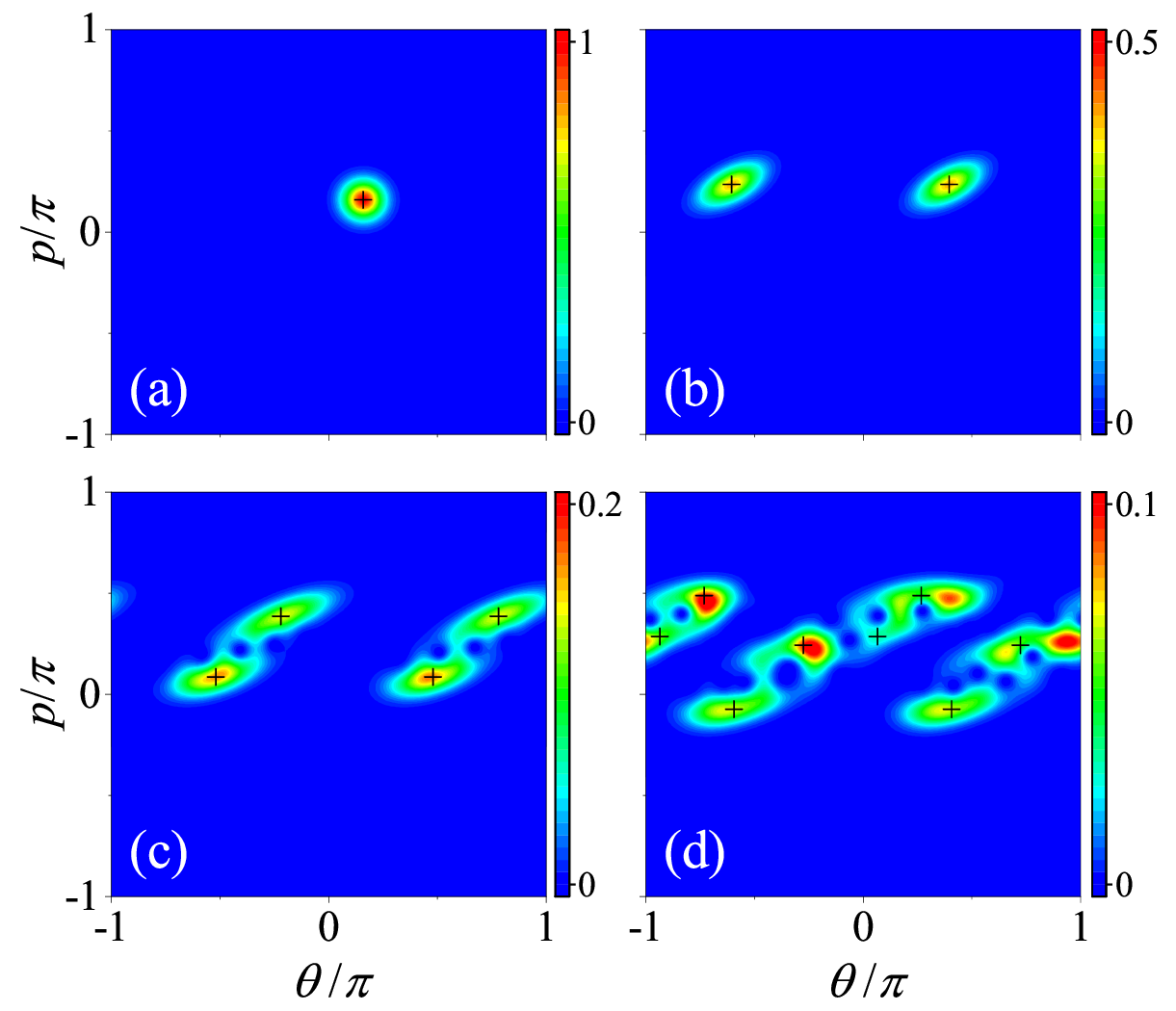}
\caption{ Husimi distribution for the quantum state at time $t = 0$ (a), $t = 1$ (b), $t = 2$ (c), and $t = 3$ (d) for a general case of $r = 1$, $s = 4$, and $\omega = 1$. Here, $\delta = 0.04$, $k = 0.5$, and the initial state is $|p_0, \theta_0\rangle=|0.5,0.5\rangle$. The black pluses denote the resultant states by the pseudoclassical map
[Eq.~\eqref{E1}].}
\label{figE1-1}
\end{figure}

To conduct a close comparison between the quantum wavepacket evolution and its pseudoclassical counterpart, it is appropriate to visualize the quantum evolution in the phase space with the Husimi distribution~\cite{Takahashi1985}. For a given quantum state $|\psi \rangle$, at the given point $(p,\theta)$ in the phase space, the Husimi distribution $\mathcal{H}(p,\theta)$ is defined as the expectation value of the density matrix $\rho = |\psi \rangle \langle \psi|$ with respect to the corresponding coherent state $|p,\theta\rangle$. That is,
\begin{eqnarray}
\mathcal{H} (p,\theta) = \langle p,\theta|\rho|p,\theta\rangle.
\end{eqnarray}

First, for the first representative example of a general case where $r = 1$, $s = 4$, and $\omega = 1$, based on Eq.~\eqref{Gl}, we have two nonzero $G_l$ and as such $\mathcal{N}=2$.
The pseudoclassical map is
\begin{eqnarray}\label{E1}
\mathcal{M}: (p, \theta) \rightarrow
		\begin{cases}
			(\tilde p, \tilde \theta);~~A_1,\\
			(\tilde p, \tilde \theta+\pi);~~A_2,\\
		\end{cases}
\end{eqnarray}
with the complex amplitudes $A_1=\frac{1+i}{2}$ and $A_2=\frac{1-i}{2}$. In addition, based on Eq.~\eqref{Md}, $\tilde p=p+k\sin\theta$ and $\tilde \theta =\theta+\tilde p$ in this case.

\begin{figure}[H]
\centering
\includegraphics[width=7.9cm]{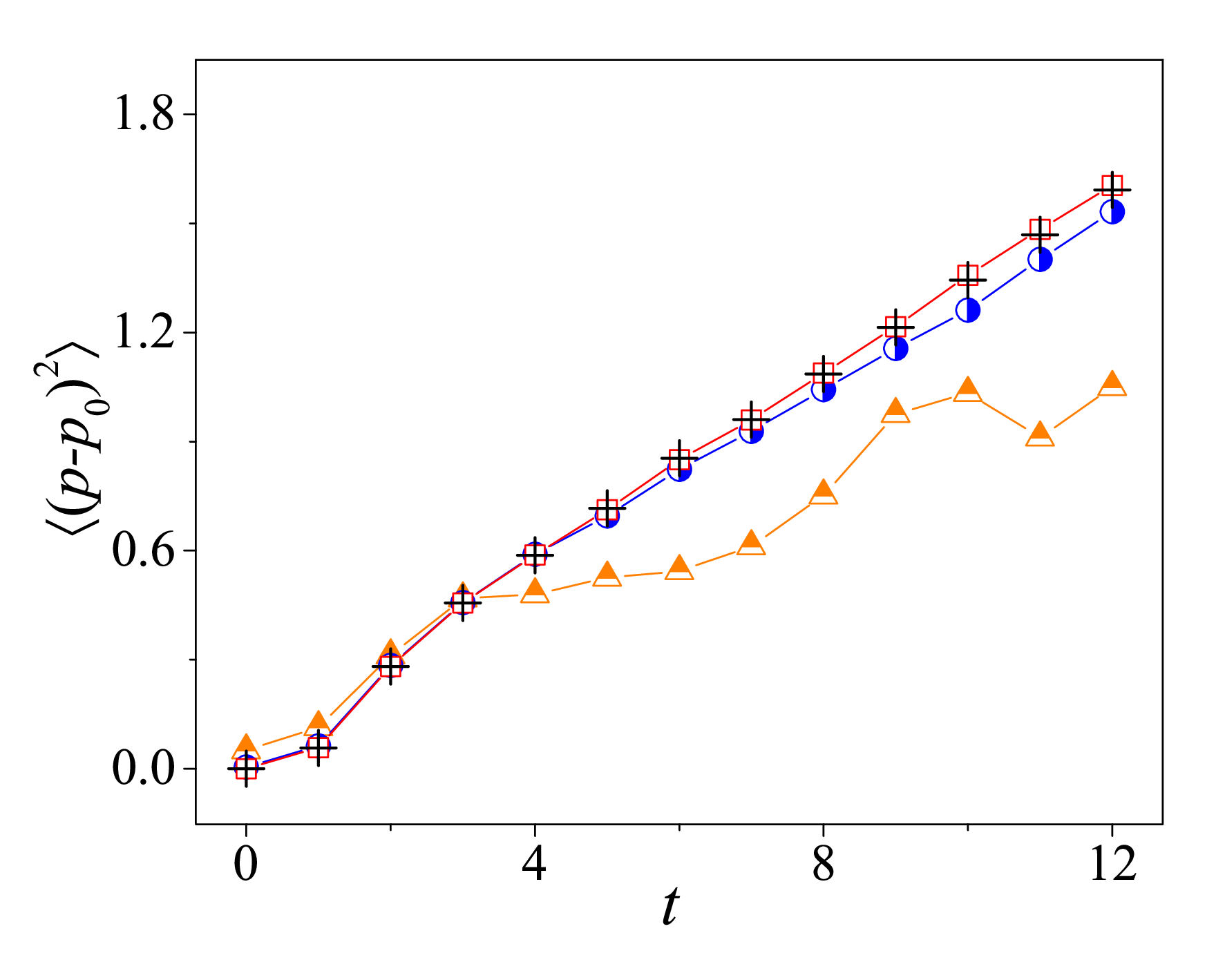}
\caption{The expected value of $(p-p_0)^2$ as a function of time for the pseudoclassical dynamics (black pluses) and the quantum dynamics with $\delta=10^{-3}$ (red squares), $10^{-2}$ (blue circles), and $10^{-1}$ (orange triangles), respectively, for the general case of $r = 1$, $s = 4$, and $\omega = 1$. Here, $k = 0.5$, and for the initial state, $p_0=\theta_0=0.5$.}
\label{figE1-2}
\end{figure}

On the quantum aspect, for a given initial coherent state $|\psi_0\rangle=|p_0, \theta_0\rangle$, the state after $t$ iterations, $|\psi_t\rangle=U^t|\psi_0\rangle$, can be numerically determined by repeatedly applying the evolution operator $U$. The contour plot of the corresponding Husimi distribution of $|\psi_t\rangle$ for $t=0$ to 3 is presented in Fig.~\ref{figE1-1}, where the wavepacket proliferation owing to quantum resonance can be clearly observed. On the pseudoclassical aspect, using the corresponding initial state $(p_0, \theta_0)$, we can acquire $\mathcal{N}^t$ states after $t$ iterations by the pseudoclassical map. The positions of these states are indicated in Fig.~\ref{figE1-1} as well. By comparison, we can observe that they can indeed well-capture the skeleton of the quantum state.

In addition to the skeleton, more information on the quantum state is encoded in the complex amplitudes of the pseudoclassical states. To determine if this information is sufficient to forecast the expected value of a given observable, the angular momentum diffusion behavior is considered as an example, quantified by $\langle (p-p_0)^2\rangle$, which is of particular interest in the quantum kicked rotor research. Fig.~\ref{figE1-2} compares the results by the quantum and the pseudoclassical dynamics. We may expect that in the limit of $\delta \to 0$, the agreement between them should be progressively enhanced, which is well corroborated.

Note that despite its success, as shown in Fig.~\ref{figE1-1} and Fig.~\ref{figE1-2}, owing to the issue of state proliferation, the implementation of the pseudoclassical theory would be prohibitively costly for a long evolution time. For instance, in the case discussed, the involved states would have reached up to $10^9$ for $t=30$.

However, for a system that has the translational symmetry in $\theta$ as in cases C1 and C2, the state proliferation challenge can be overcome. As two examples of case C1 and C2, we consider $r=1$, $s=3$, and $\omega=3$ and $r=1$, $s=4$, and $\omega=2$, respectively.

\begin{figure}[H]
{\includegraphics[width=7.9cm]{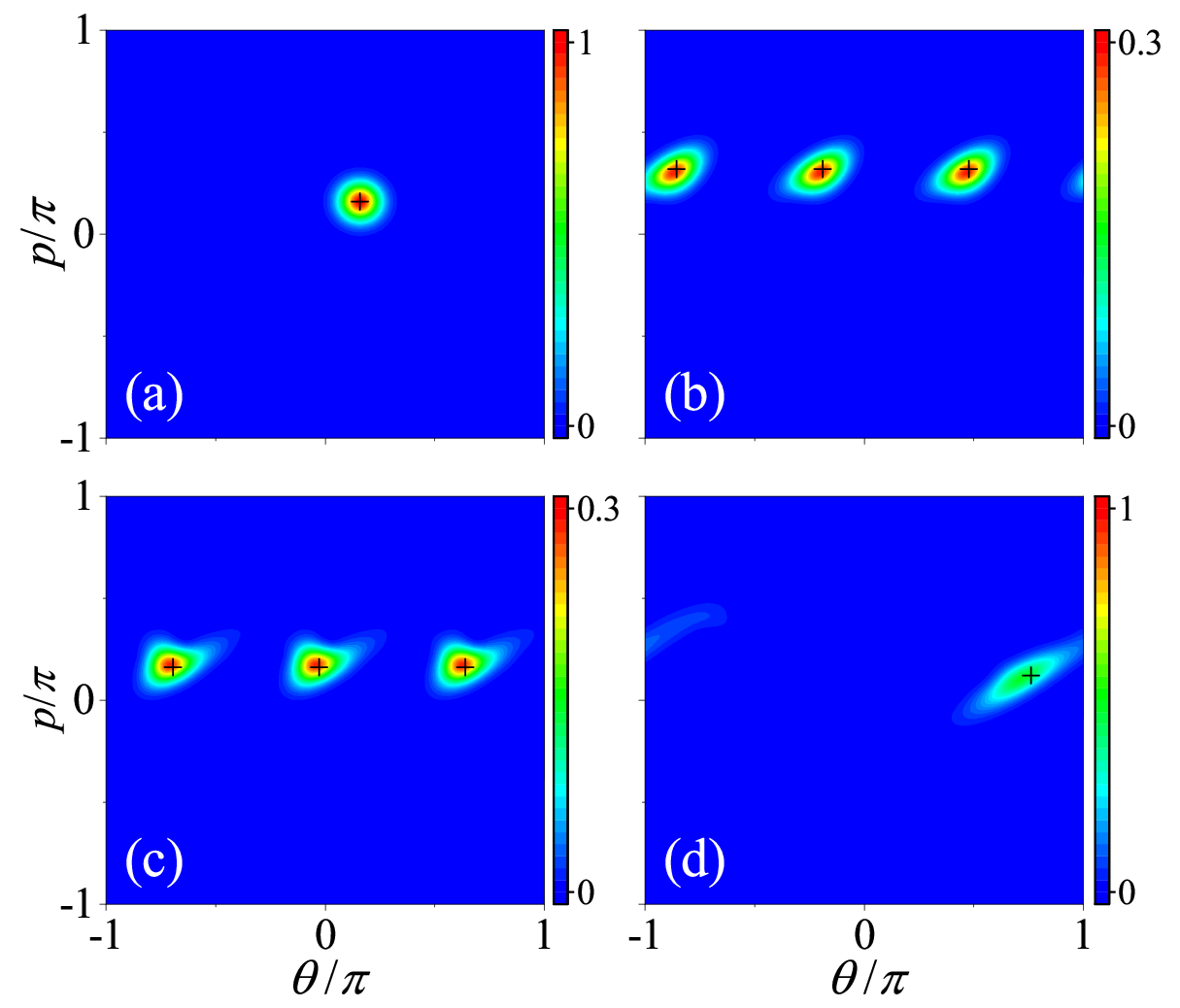}}
\caption{The same as Fig.~\ref{figE1-1} but for case C1 with $r=1$, $s=3$, and $\omega=3$.}
\label{figE2}
\end{figure}

For the former, $\mathcal{N}=3$ and the pseudoclassical map reads as
\begin{eqnarray}\label{E2}
\mathcal{M}: (p, \theta) \rightarrow
		\begin{cases}
			(\tilde p, \tilde \theta);~~A_1,\\
			(\tilde p, \tilde \theta+\frac{2 \pi}{3});~~A_2,\\
            (\tilde p, \tilde \theta+\frac{4 \pi}{3});~~A_3,\\
		\end{cases}
\end{eqnarray}
where $A_1=-\frac{\sqrt{3}i}{3}$, $A_2=A_3=\frac{3+\sqrt{3}i}{6}$, and $(\tilde p,\tilde \theta)$ is generated from $(p,\theta)$ according to Eq.~\eqref{Md}. This suggests that, after each step, a point will be mapped into three of the same weight but of two different phases.

\begin{figure}[H]
{\includegraphics[width=7.9cm]{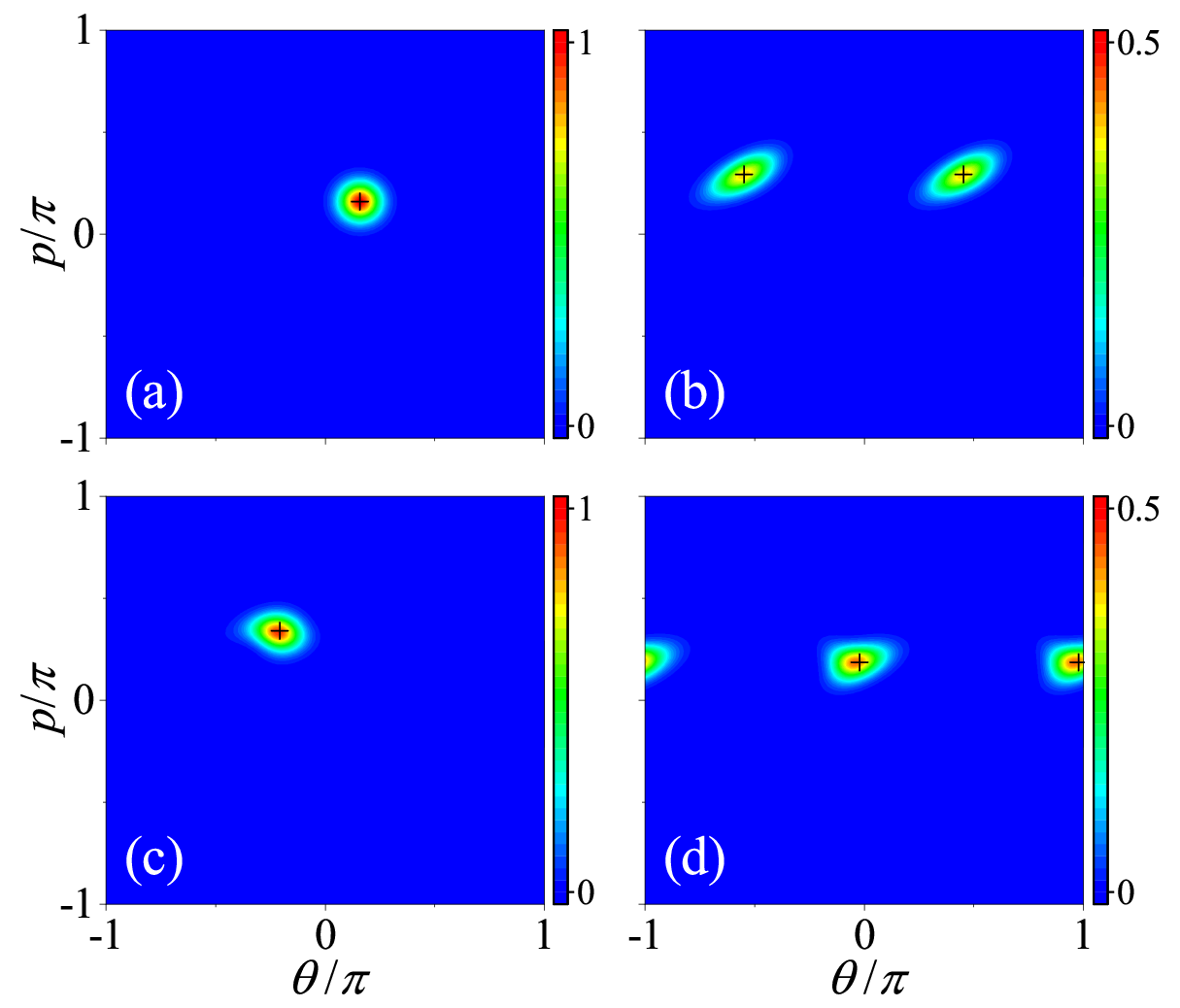}}
\caption{The same as Fig.~\ref{figE1-1} but for case C2 with $r=1$, $s=4$, and $\omega=2$. }
\label{figE3}
\end{figure}


The results of the Husimi distribution of the quantum state at $t = 0$ to 3 and the pseudoclassical map are shown in Fig.~\ref{figE2}. It can be observed that at $t=1$, the initial quantum coherent state is mapped into three, and their centers perfectly overlap with the three resulting pseudoclassical states. At $t=2$, of the nine expected pseudoclassical states, three pairs cancel each other so that only three survive. This is completely supported by quantum evolution. Finally, at $t=3$, the three remaining pseudoclassical states merge into one rather than split into nine, successfully further predicting quantum evolution. For the quantum evolution, it can be straightforwardly shown that $U_f^3=1$, suggesting that the number of involved coherent states must reduce to one after every three steps.

For the latter case of $r=1$, $s=4$, and $\omega=2$, the pseudoclassical map is the same as Eq.~\eqref{E1} but with $\tilde p=p+k\sin(2\theta)$ and $\tilde \theta =\theta+\tilde p$ instead. As $A_{1,2} =\frac{1\pm i}{2}$, a point will be mapped into two with the same weight but different phases. Fig.~\ref{figE3} compares the quantum and pseudoclassical dynamics, and good agreement between them is the same as in the previous case. Note that for this case, although $U_f^4 =1$, $U_f^2 \ne 1$. However, ${U'}_f^2=1$ such that $(U_f U_\delta)^2={U'}_\delta^2$, which explains why after every two steps, the number of pseudoclassical states becomes one.

\begin{figure}[H]
\centering
\includegraphics[width=7.9cm]{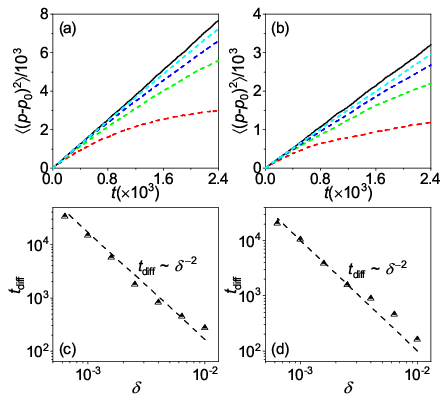}
\caption{(a) Time dependence of $\langle (p-p_0)^2 \rangle$ for $r=1$, $s=3$, and $\omega=3$. Red, green, blue, and cyan dashed curves are for the quantum results with $\delta= 10^{-2}$, $10^{-2.4}$, $10^{-2.8}$, and $10^{-3.2}$, respectively, for the initial state $|\psi(0)\rangle = |0\rangle$. The black curve is for the ensemble average of the pseudoclassical results for $10^6$ initial states uniformly distributed on the line of $p=0$ as the classical counterpart of $|0\rangle$. (b) Same as (a) but for $r=1$, $s=4$, and $\omega=2$. Panels (c) and (d) are for the $\delta$ dependence of the diffusion time $t_{\text{diff}}$ for the quantum system presented in (a) and (b), respectively. Here, in all the simulations, $k=2$.}
\label{E23-p2}
\end{figure}

In these two cases, as the state proliferation problem is well suppressed owing to translational symmetry, the pseudoclassical evolution can be conveniently conducted up to a much longer time than in the general case. For these two cases, Figs.~\ref{E23-p2}(a) and \ref{E23-p2}(b) compare the time dependence of $\langle (p-p_0)^2 \rangle$ calculated with the quantum and the pseudoclassical dynamics over a wide time range. In particular, $k$ is fixed, but perturbation $\delta$ is changed to observe how the quantum results depend on it. Indeed, as expected, as $\delta$ decreases, the quantum result tends to approach the pseudoclassical result. In addition, for the diffusion time $t_{\text{diff}}$, which is empirically defined as the time when the quantum result deviates from the pseudoclassical result by $15\%$ from below, it follows the scaling $t_{\text{diff}} \sim \delta^{-2}$ [see Figs. ~\ref{E23-p2}(c) and \ref{E23-p2}(d)], the same as in the conventional semiclassical limit of the kicked rotor model if $\delta$ is recognized with the effective Planck constant~\cite{Izrailev1990}. All these results consistently support the effectiveness of our pseudoclassical theory.

\section{Application to the $\mathcal{PT}$-symmetric kicked rotor}

Quantum mechanical Hamiltonians that are $\mathcal{PT}$-symmetric but not Hermitian has recently been a frontier subject ~\cite{Bender1998, Graefe2010, West2010, Bender2015, Longhi2017, Ganainy2018, Zhao2019}. A Hamiltonian $H$ is regarded as $\mathcal{PT}$-symmetric if $\lbrack H, \mathcal{PT}\rbrack = 0$, where the parity operator, $\mathcal{P}$, is a unitary operator that satisfies $\mathcal{P}^2=1$ and the time-reversal operator, $\mathcal{T}$, is an antiunitary operator that satisfies $\mathcal{T}^2=\pm 1$. As a result, $\mathcal{PT}$ is an antiunitary operator as well. Surprisingly, as observed in some previous works and highlighted in Ref.~\cite{Bender1998}, it is feasible for a $\mathcal{PT}$-symmetric Hamiltonian to have a real spectrum, despite the fact that it can be non-Hermitian. Moreover, as the gain (or loss) parameter $\lambda$ that controls the degree of non-Hermiticity changes, a spontaneous $\mathcal{PT}$ symmetry breaking may take place.

\begin{figure}[H]
{\includegraphics[width=7.9cm]{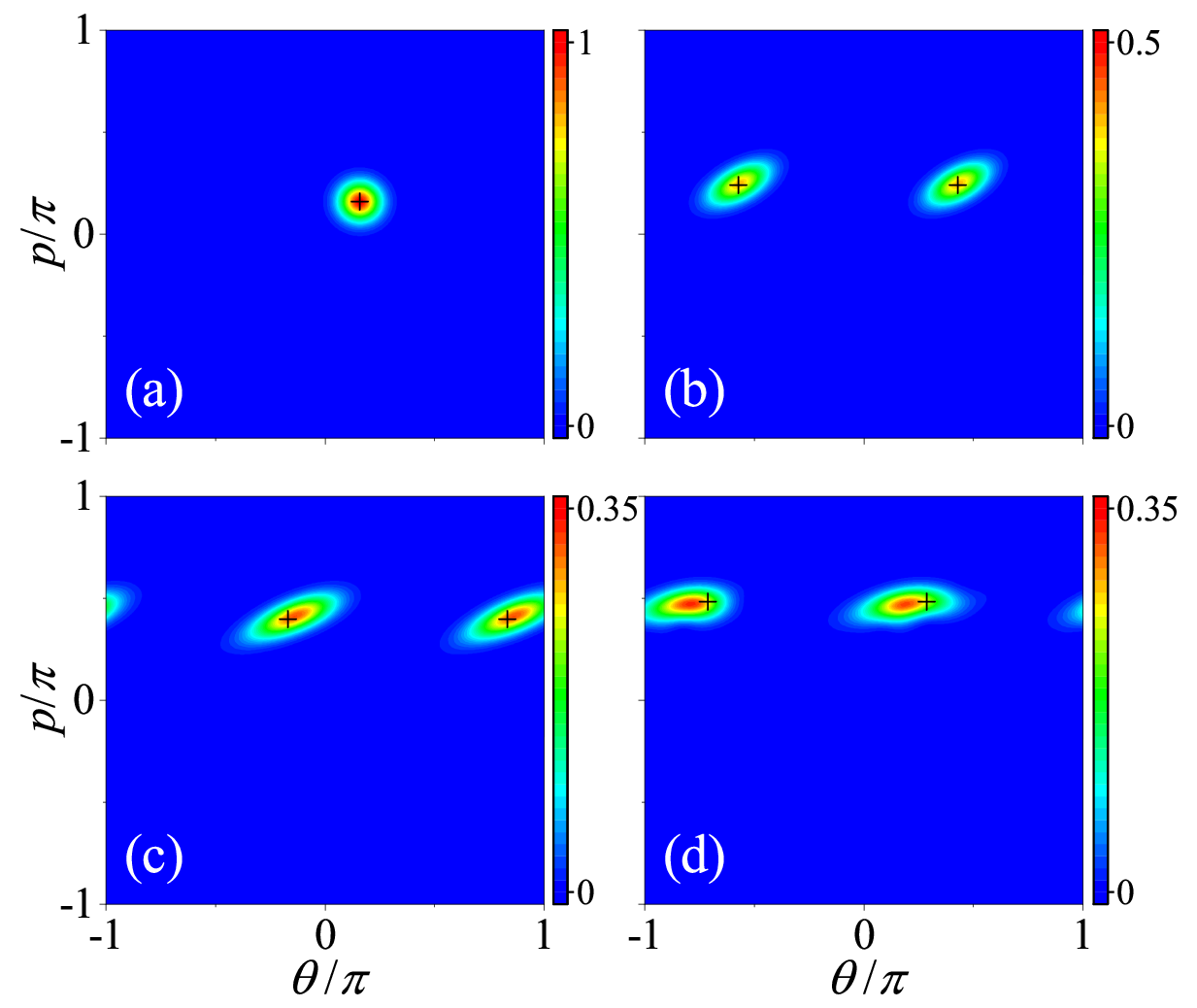}}
\caption{The same as Fig.~1 but for the $\mathcal{PT}$-symmetric kicked model with
$\lambda=0.2$. (The results shown in Fig.~1 are equivalent to the case of $\lambda=0$.)
The black pluses stand for the positions of the resultant states generated by the
pseudoclassical map $\mathcal{M}^{\mathrm{NH}}$ [see Eq. \eqref{NH-M}].}
\label{figE6}
\end{figure}

In this section, we attempt to use our pseudoclassical theory to the $\mathcal{PT}$-symmetric kicked rotor model whose Hamiltonian is~\cite{Longhi2017}
\begin{eqnarray}\label{PT-ham}
  H^\mathrm{NH}=\frac{P^2}{2I}+\frac{KI}{\omega T}(\cos \omega\theta +
  i\lambda \sin\omega\theta)\sum_{m=-\infty}^{\infty} \delta(\tau-mT),
\end{eqnarray}
where $\lambda\ge 0$ is the non-Hermitian parameter that controls the strength of the imaginary part of the kicking potential. For $\lambda = 0$, it reduces to the conventional Hermitian kicked rotor. When $\alpha\equiv\hbar T/I=4\pi\frac{r}{s}$ with $r$ and $s$ two coprime integers, similar to the conventional kicked rotor, the quasienergy spectrum is absolutely continuous and composed of $s$ quasienergy bands, which is also named quantum resonance~\cite{Longhi2017}. In the following, we will focus on the perturbed case where $\alpha=\frac{4\pi r}{s}+\delta$, for which the Floquet operator can be written as
\begin{eqnarray}\label{NH-U}
U^\mathrm{NH}&=&\exp\left(-i2\pi\frac{r}{s}\hat \nu^2\right)\exp\left(-i\frac{\delta}{2}
\hat \nu^2\right)\nonumber\\
&~&\times \exp(-i\frac{k}{\delta}\cos(\omega\theta))\exp(\frac{k\lambda}{\delta }\sin(\omega\theta)).
\end{eqnarray}
It includes two parts: $U^\mathrm{NH}=U_f U_\delta^\mathrm{NH}$ with $U_f=\exp\left(-i2\pi\frac{r}{s}\hat \nu^2\right)$ and $U_\delta^\mathrm{NH}= \exp\left(-i\frac{\delta}{2}\hat \nu^2\right) \exp(-i\frac{k}{\delta}\cos(\omega\theta)) \\\exp(\frac{k\lambda}{\delta }\sin(\omega\theta))$.
The difference lies in the last term of $U_\delta^\mathrm{NH}$. For $\lambda=0$, this term is identity, $U_\delta^\mathrm{NH}$ reduces to $U_\delta$ and $U^\mathrm{NH}$ reduces to $U$.

For $\lambda\ne 0$ and in the limit of $\delta\to 0$, when the last term of $U_\delta^{\mathrm{NH}}$ acts on a quantum state expressed as the superposition of multiple coherent states, only the component coherent state $|p_j, \theta_j\rangle$ that maximizes $\sin(\omega\theta)$ (with $\theta_j$) significantly contributes to the result; the contributions of other components can be ignored. In such a sense, one role that the last term of $U_\delta^{\mathrm{NH}}$ plays is a ``selector''. Formally, the pseudoclassical counterpart of this role can be denoted as $\mathcal{M}_s$. It chooses the component state $(p_j, \theta_j)$ from others as the first step of the pseudoclassical dynamics.

\begin{figure}[H]
{\includegraphics[width=7.9cm]{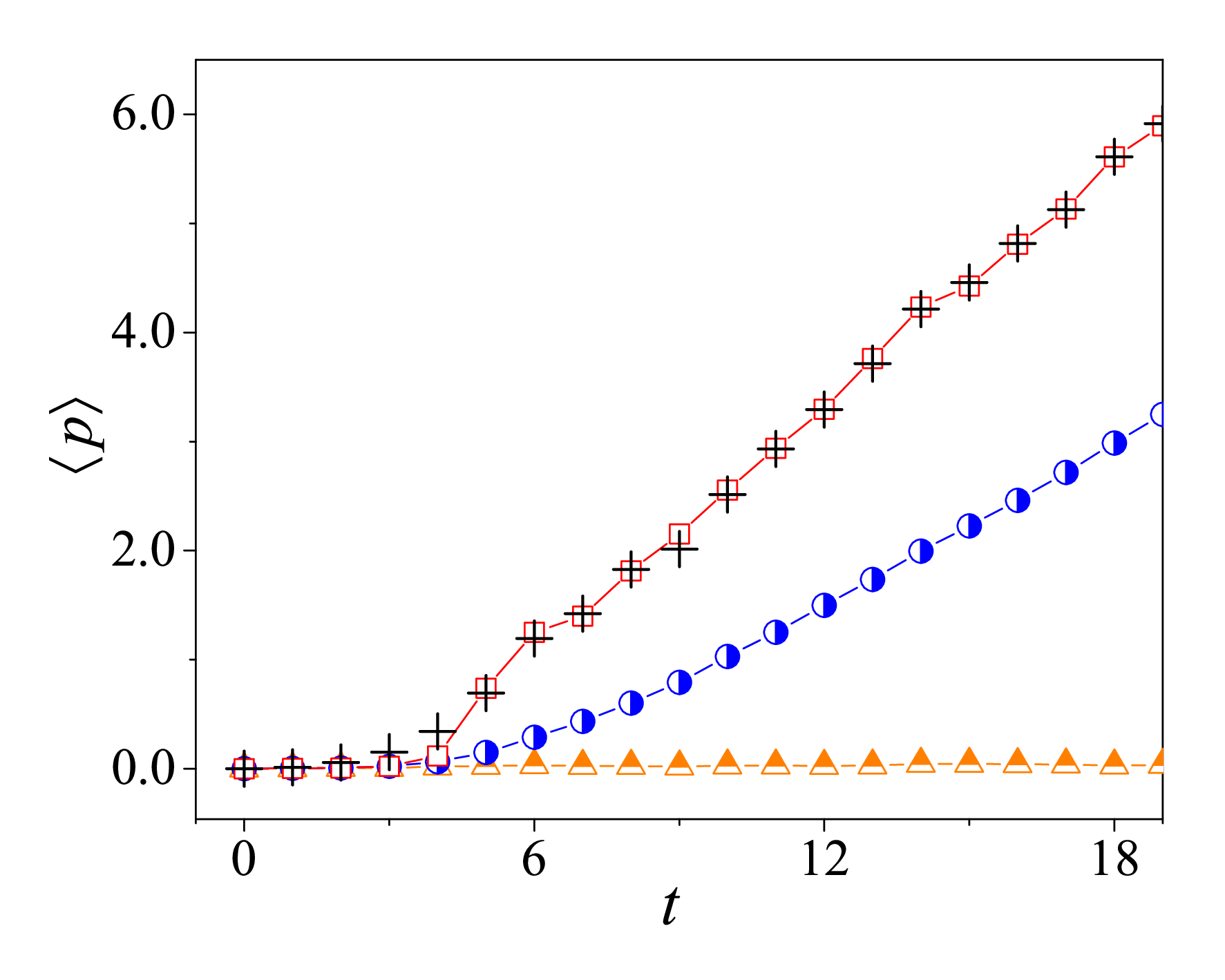}}
\caption{ Expected value of the momentum as a function of time for the pseudoclassical dynamics (black pluses) and the quantum dynamics with $\delta =10^{-3}$ (red squares), $10^{-2}$ (blue circles), and $10^{-1}$ (orange triangles), respectively, of the $\mathcal{PT}$-symmetric kicked model with $\lambda=0.01$. Here, $r = 1$, $s = 4$, $\omega = 1$, $k = 0.5$, and for the initial state, $p_0 = \theta_0 = 0$.}
\label{figE7}
\end{figure}

The last term of $U_\delta^{\mathrm{NH}}$ also plays a key role in retrieving the pseudoclassical counterpart, denoted as $\mathcal{M}_\delta^{\mathrm{NH}}$, of $U_\delta^{\mathrm{NH}}$. Based on the generalized canonical structure theory \cite{Graefe2010}, in the limit of $\delta\to 0$, the motion of the center of a given coherent state $U_\delta^{\mathrm{NH}}$ acts on is governed by the following equations:
\begin{eqnarray}\label{NH_cl}
     \dot{p} &=& k\sin\omega\theta\sum_{m=-\infty}^{\infty} \delta(\tau-m) ,\nonumber\\
     \dot{\theta} &=& p+k\lambda\cos\omega\theta\sum_{m=-\infty}^{\infty} \delta(\tau-m).
\end{eqnarray}
Thus, the pseudoclassical operation $\mathcal{M}_\delta^{\mathrm{NH}}$ thus represents the integration of these two functions up to a unit time. For $\lambda=0$, it reduces to $\mathcal{M}_\delta$ given by Eq.~\eqref{Md}.

Formally, the pseudoclassical map for the $\mathcal{PT}$-symmetric kicked rotor can be written as
\begin{eqnarray}\label{NH-M}
\mathcal{M}^{\mathrm{NH}}=\mathcal{M}_f\mathcal{M}^{\mathrm{NH}}_\delta\mathcal{M}_s.
\end{eqnarray}
To test its effectiveness, a general case of the $\mathcal{PT}$-symmetric kicked rotor is simulated with $r = 1$, $s = 4$, and $\omega = 1$, the same as in Figs.~1 and 2. Figs.~6 and 7 illustrate the results. Fig.~6 displays the Husimi distribution of the quantum state evolved from an initial coherent state. Owing to the gain (or loss) operation of the selecting operator $\exp(\frac{k\lambda}{\delta } \sin\theta)$, only two component coherent states appear for $t\ge 2$, in clear contrast to the conventional kicked rotor corresponding to $\lambda=0$ (see Fig.~1 for comparison). Namely, the state proliferation problem in the latter is effectively suppressed here by the selection operator. Meanwhile, we can see that the positions of the two component coherent states are well predicted by the pseudoclassical map.

An interesting feature of the $\mathcal{PT}$-symmetric kicked rotor is that it can generate the directed current~\cite{Longhi2017, Zhao2019}. Here we study this property with $\alpha$ being slightly perturbed from the quantum resonance condition. Fig.~\ref{figE7} compares the expected values of the momentum, $\langle p \rangle$, for quantum and pseudoclassical dynamics. It can be observed that as $\delta$ decreases, they do converge and $\langle p \rangle\sim t$, which implies that the directed current also exists in the deep quantum regime near quantum resonances and interestingly, it has a pseudoclassical explanation.

\section{Summary}

In this study, by considering the quantum effect of the free rotation at quantum resonances, a generalized pseudoclassical theory is designed for the kicked rotor model. Its effectiveness suggests that, even in the deep quantum regime, quantum dynamics may have a close connection to classical dynamics. With regard to this, one may wonder if it is imperative to extend the conventional quantum chaos study from the semiclassical regime~\cite{Berry89} to the deep quantum regime. In this context, it depends on how general the pseudoclassical theory could be, which warrants further investigation, except the kicked top and the kicked rotor model, if the pseudoclassical theory can be extended to other Floquet systems.

Extensive experiments have been conducted on the kicked rotor model with cold atoms~\cite{Garreau2020} owing to its paradigmatic role in illustrating quantum chaos. It would be enticing to examine the effects of the pseudoclassical dynamics near higher-order quantum resonances. To this end, it is necessary to adapt the pseudoclassical theory to the kicked particle model first, which is in progress.

\Acknowledgements{This work is supported by the National Natural Science Foundation of China
(Grants No. 12075198, No. 12247106, and No. 12247101).}

\InterestConflict{The authors declare that they have no conflict of interest.}



\end{multicols}
\begin{appendix}
\section{Derivation of Equation~\eqref{Uf1} }
Regarding the eigenstates $\lbrace |n \rangle\rbrace$ of the operator $\hat\nu$, the coherent state $|{\tilde p, \tilde \theta}\rangle $ can be expressed as $|{\tilde p, \tilde \theta}\rangle= \sum_{n} c_n|n\rangle$. Applying the operator $\exp\left( -i\frac{2\pi r}{s}\hat\nu^2\right)$ to both sides,
\begin{eqnarray}\label{eqnA1}
\exp\left(-i\frac{2\pi r}{s}\hat{\nu}^2\right)|{\tilde p, \tilde \theta}\rangle=\sum_{k=0}^{s-1}
\exp\left(-i\frac{2\pi r}{s}k^2\right) \sum_{\mod(n,s)=k} c_n|n\rangle.
\end{eqnarray}
Note that two coherent states separating in $\theta$ by $\phi$ can be related by the translation operator $\exp(-i\hat{\nu}\phi)$, we have
\begin{eqnarray}
|\tilde p, \tilde \theta+\frac{2\pi r}{s}l\rangle &=& \exp\left(-i\hat{\nu}
\frac{2\pi r}{s}l\right)|\tilde p, \tilde \theta\rangle\nonumber\\
&=& \sum_{k=0}^{s-1} \exp\left(-i\frac{2\pi kr}{s}l\right)\sum_{\mod(n,s)=k}c_n|n\rangle
\end{eqnarray}
by setting $\phi=2\pi lr/s$, where $l$ is an integer. Thereafter, multiplying both sides with $\exp(i2\pi \lambda lr/s)$, where $\lambda$ is an integer, $0\le \lambda \le s-1$, and taking summation over $l$ from $l=0 $ to $s-1$,
\begin{eqnarray}
\sum_{l=0}^{s-1}\frac{1}{s}\exp\left(i\frac{2\pi r}{s}\lambda l\right)|p_c,\theta_c+
\frac{2\pi r}{s}l\rangle =\sum_{\mod(n,s)=\lambda} c_n |n\rangle.
\end{eqnarray}
Finally, by replacing $\lambda$ with $k$ and substituting this equation into Eq.~~\eqref{eqnA1}, we get Eq.~\eqref{Uf1}.

\section{ Translational symmetry of cases C1 and C2}
To prevent the proliferation problem of the pseudoclassical dynamics, one way is to introduce the translational symmetry into the phase space so that two or more state points will be mapped into one by $\mathcal{M}_f$. This can be fulfilled by setting the proper integer value of $\omega$.

First, for case C1 where $s$ is odd, based on the analytical results by Ref.~\cite{Berndt1998}, all $s$ Gaussian sums are nonzero so that $\mathcal{N}=s$ and any two neighboring state points resulted by acting $\mathcal{M}_f$ to a given state are separated in $\theta$ by $2\pi/s$. Hence, to ensure the resultant state points overlap at the next step, we can set $\omega=s$. It ensures that the number of states is up-bounded by $s$ throughout.

For case C2, where $s$ is even, note that the Gaussian sum $G_l$ can be rewritten as
\begin{eqnarray}
G_l &=& \frac{1}{s} \sum_{k=0}^{s-1} \exp\left(-i\frac{2\pi r}{s}k(k-l)\right)\nonumber\\
&=& \frac{1}{s} \sum_{k=0}^{s/2-1}\exp\left(-i\frac{2\pi r}{s}k(k-l)\right)+\exp\left(-i\frac{2\pi r}{s}(k+\frac{s}{2})(k+\frac{s}{2}-l)\right)\nonumber\\
&=& \frac{1}{s} \sum_{k=0}^{s/2-1}\exp\left(-i\frac{2\pi r}{s}k(k-l)\right)
\left(1+\exp\left(-i2\pi r\left(\frac{s}{4} -\frac{l}{2}\right)\right)\right).
\end{eqnarray}
It is evident that the last term, $1+\exp\left(-i2\pi r\left(\frac{s}{4}-\frac{l}{2}\right)\right)$, is zero for all odd $l$ when mod$(s,4)=0$ and is zero for all even $l$ when mod$(s,4)=2$. On the one hand, this implies that half of all $s$ Gaussian sums are zero, whereas the other half are nonzero based on Ref.~\cite{Berndt1998}, so that $\mathcal{N}= s/2$. On the other hand, this suggests that of all $s/2$ state points resulted by acting $\mathcal{M}_f$ to a given state, any two neighboring points are separated in $\theta$ by $4\pi/s$. Thus, similar to case C1, to ensure the evolving state points overlap at the following steps, we can set $\omega=\mathcal{N}=s/2$. This guarantees that the number of states is up-bounded by $\mathcal{N}=s/2$ throughout.

\section{Commutation relation of $U_f$ and $U_\delta$}

In the representation of $\hat{\nu}$, the elements of $U_f$ and $U_\delta$ are, respectively,
\begin{eqnarray}
  (U_f)_{m,m'} &=& \exp\left(-i\frac{2\pi r}{s} m^2\right)\delta_{m,m'},\\
  (U_\delta)_{m,m'} &=& \frac{1}{2\pi}\exp\left(-i\frac{m^2}{2}\delta\right)
  \int_0^{2\pi} \exp\left(-i(m-m')\theta\right)
  \exp\left(-i\frac{k}{\omega\delta}\cos\left(\omega\theta
  \right)\right) d\theta,
\end{eqnarray}
following which the matrix element for the commutator $( U_f U_\delta-U_\delta U_f)_{m,m'}$ is
 \begin{eqnarray}\label{commutator}
   ( U_f U_\delta-U_\delta U_f)_{m,m'} &=& \sum_n (U_f)_{m,n}(U_\delta)_{n,m'}
   - \sum_n (U_\delta)_{m,n}(U_f)_{n,m'} \nonumber\\
   &=& ((U_f)_{mm}-(U_f)_{m',m'}) (U_\delta)_{m,m'}\nonumber\\
   &=&\left( \exp\left(-i\frac{2\pi r}{s}m^2\right)-\exp\left(-i\frac{2\pi r}{s}m'^2\right)\right)
   \frac{\exp\left(-i\frac{m^2}{2}\delta\right)}{2\pi}\int_0^{2\pi} \exp\left(-i(m-m')\theta\right)
   \exp\left(-i\frac{k}{\omega\delta}\cos\left(\omega\theta\right)\right) d\theta.
 \end{eqnarray}
Note that $\exp\left(-i\frac{k}{\omega\delta}\cos\left(\omega\theta\right)\right)$ is periodic with the period $2\pi/\omega$; thus, the integration term in the above equation can be rewritten as
 \begin{eqnarray}\label{sumterm}
\int_0^{2\pi} \exp\left(-i(m-m')\theta\right)\exp\left(-i\frac{k}{\omega\delta}
   \cos\left(\omega\theta\right)\right) d\theta  &=&\int_0^{\frac{2\pi}{\omega}}\left(\sum_{u=0}^{\omega-1}\exp\left(-i(m-m')\left(\theta+\frac{2\pi u}{\omega}\right)\right)\right) \exp\left(-i\frac{k}{\omega\delta}\cos\left(\omega\theta\right)\right) d\theta\nonumber\\
   &=&\int_0^{\frac{2\pi}{\omega}}\left(\exp\left(-i(m-m')\theta\right)\sum_{u=0}^{\omega-1}\exp\left(-i\frac{2\pi (m-m') u}{\omega}\right)\right) \exp\left(-i\frac{k}{\omega\delta}\cos\left(\omega\theta\right)\right) d\theta \nonumber\\
   &=&\omega\delta_{\mathrm{mod}(m-m',\omega),0}\int_0^{\frac{2\pi}{\omega}}\exp\left(-i(m-m')\theta\right)\exp\left(-i\frac{k}{\omega\delta}\cos\left(\omega\theta\right)\right) d\theta .
 \end{eqnarray}
 Substituting it into Eq. \eqref{commutator},
  \begin{eqnarray}\label{C1commutator}
   ( U_f U_\delta-U_\delta U_f)_{m,m'} &=& \left( \exp\left(-i\frac{2\pi r}{s}m^2\right)-\exp\left(-i\frac{2\pi r}{s}m'^2\right)\right)\times \frac{\omega}{2\pi}\exp\left(-i\frac{m^2}{2}\delta\right)\nonumber\\
   &~&\delta_{\mathrm{mod}(m-m',\omega),0}\int_0^{\frac{2\pi}{\omega}}\exp\left(-i(m-m')\theta\right)\exp\left(-i\frac{k}{\omega\delta}\cos\left(\omega\theta\right)\right) d\theta.
 \end{eqnarray}
For case C1 where $s$ is odd, as $\omega=s$, the delta function in Eq.~\eqref{C1commutator} suggests that $m-m'=ns$ ($n$ is an integer). Substituting it into Eq.~\eqref{commutator}, it becomes
\begin{eqnarray}
  ( U_f U_\delta-U_\delta U_f)_{m,m'} =0
\end{eqnarray}
because the difference term in Eq.~\eqref{commutator} is
 \begin{eqnarray}
\left( \exp\left(-i\frac{2\pi r}{s}m^2\right)-\exp\left(-i\frac{2\pi r}{s}m'^2\right)\right)&=& \exp\left(-i\frac{2\pi r}{s}m^2\right)\left( 1-\exp\left(-i2\pi r(n^2s-2mn)\right)\right)\nonumber\\
& =& 0.
 \end{eqnarray}

However, for case C2 where $s$ is even and $\omega=\frac{s}{2}$, the difference term in Eq.~\eqref{commutator} does not equal zero in general, and thus, $U_f U_\delta-U_\delta U_f\ne 0$. However, if we split the Floquet operator $U$ as $U=U'_f U'_\delta$ with $U'_f = U_f \exp(i \pi r \nu)$ and $U'_\delta = \exp(-i\pi r \nu)U_\delta$, as shown in the following, $U'_f$ and $U'_\delta$ do commutate.

The matrix element for $U'_f$ and $U'_\delta$ is, respectively,
\begin{eqnarray}
  (U'_f)_{m,m'} =(U_f)_{m,m'}\exp(i\pi r m') {\rm ~~and~~ } (U'_\delta)_{m,m'} =\exp(-i\pi r m)(U_\delta)_{m,m'},
\end{eqnarray}
and thus,
 \begin{eqnarray}\label{commutator2}
   ( U'_f U'_\delta-U'_\delta U'_f)_{m,m'} &=& \sum_n (U'_f)_{m,n}(U'_\delta)_{n,m'} - \sum_n (U'_\delta)_{m,n}(U'_f)_{n,m'} \nonumber\\
   &=& ((U'_f)_{mm}-(U'_f)_{m',m'}) (U'_\delta)_{m,m'}\nonumber\\
   &=&\left( \exp\left(-i\frac{2\pi r}{s}m^2+i\pi rm\right)-\exp\left(-i\frac{2\pi r}{s}m'^2+i\pi rm' \right)\right)\frac{1}{2\pi}\exp\left(-i\frac{m^2}{2}\delta-i\pi rm\right) \nonumber\\ &~&\times\int_0^{2\pi}\exp\left(-i(m-m')\theta\right)\exp\left(-i\frac{k}{\omega\delta}\cos\left(\omega\theta\right)\right) d\theta.
 \end{eqnarray}
Note that here the integration term is the same as Eq.~\eqref{commutator}; however, the delta function suggests $m-m'=n \frac{s}{2}$ instead ($n$ is an integer). Similarly, substituting it into Eq.~\eqref{commutator2}, we obtain
\begin{eqnarray}
( U'_f U'_\delta-U'_\delta U'_f)_{m,m'}=0,
\end{eqnarray}
because the difference term in Eq.~\eqref{commutator2} is
\begin{eqnarray}
\exp\left(-i\frac{2\pi r}{s}m^2+i\pi rm\right)-\exp\left(-i\frac{2\pi r}{s}m'^2+i\pi rm' \right)&=&\exp\left(-i\frac{2\pi r}{s}m^2+i\pi rm\right)\left(1-\exp\left(-i2\pi r\left(mn - \frac{n(n-1)s}{4}\right)\right)\right)\nonumber\\
&=& 0
\end{eqnarray}
for any even $s$.

\end{appendix}

\end{document}